\documentstyle[preprint,aps,prl]{revtex}
\begin{document}
\draft
\title{ Optimized effective potential method 
with exact exchange and static RPA correlation} 
\author{Takao Kotani}
\address{Department of Physics,
Osaka University, Toyonaka 560, Japan}
\date{\today}
\maketitle
\begin{abstract}
     We present a new density-functional method 
of the self-consistent electronic-structure calculation 
which does not exploit any local density approximations (LDA).
We use the exchange-correlation energy which consists
of the exact exchange and the correlation energies in 
the random-phase approximation. 
The functional derivative of the correlation energy
with respect to the density is obtained
within a static approximation.
For transition metals, it is shown that 
the correlation potential gives rise to a large 
contribution which has the opposite sign to the exchange potential.
Resulting eigenvalue dispersions and the magnetic moments 
are very close to those of LDA's and the experiments.
\end{abstract}

\narrowtext
The optimized-effective-potential (OEP) method
for the electronic-structure calculations was first applied to atoms
by Talman and Shadwick \cite{talman76}, who recognized it as a kind of 
restricted minimum search for the Hartree-Fock total energy. 
In the method, the variational 
space was restricted to the space of the local one-particle potential, 
which generates all the eigenfunctions, or 
to the space of the corresponding density. 
From the view of the density-functional (DF) theory, 
their calculations are recognized as the Kohn-Sham (KS) 
exact exchange (EXX) -only DF calculations. 
In the last years, we have extended their method to
be applicable for solids \cite{kotani94}-\cite{kotani96kkr}, where 
we add the correlation energy in the local density approximation (LDA).
As was shown there, 
the results admittedly were not satisfactory 
from the view point of comparison with experiments.
Especially for transition metals such as iron \cite{kotani96fe},
the method gave the occupied $d$ bands which were too deep
relative to the $s$ bands and also gave rise to too large magnetizations.
This indicates that the LDA correlation is poor when we combine it 
with EXX, and that the true correlation should give rather large contributions
cancelling large EXX.

There are two kinds of way to take into account the correlation energy
more precisely. One is the multiple-configuration extension
\cite{aashamar79}, which is a widely-used idea in the quantum chemistry.
The other is based on the DF theory, to utilize the implicit DF 
for the correlation energy based on the KS orbitals, 
in addition to the EXX energy. 
Along the second, Garbo and Gross tried a new DF for
correlation, which applied to atoms and molecules \cite{garbo95}.

In this Letter, we present a new method for solid along the second line.
We use an exchange-correlation (XC) energy functional, which 
consists of the EXX energy and a correlation energy 
in the random-phase approximation (RPA).
The correlation potential, which is the functional derivative of the
correlation energy with respect to the density, 
is evaluated within a static approximation;
the static screened-Coulomb interaction,
which is used for the evaluation of the RPA correlation, 
is calculated by use of the product-basis 
method developed by Aryasetiawan and Gunnarson \cite{aryasetiawan94}.
We denote our method as EXX+RPA in the following.

Formally, the XC energy $E_{\rm xc}[n]$ as a DF is
given by use of the coupling-integral method:
\begin{eqnarray}
&&E_{\rm xc}[n] =
   \frac{e^2}{2} \sum_{\sigma,{\sigma'}}
   \int_0^1 d\lambda 
   \int d{\bf r}_1 d{\bf r}_2
    \frac{\langle 
            \hat{\psi}^{\dagger}_{\sigma}({\bf r}_1)
            \hat{\psi}^{\dagger}_{{\sigma'}}({\bf r}_2)
            \hat{\psi}_{{\sigma'}}({\bf r}_2)
            \hat{\psi}_{\sigma}({\bf r}_1)   
        \rangle_{n,\lambda}-n_{\sigma}({\bf r}_1) n_{\sigma'}({\bf r}_2)}
   {|{\bf r}_1-{\bf r}_2|}, 
\label{exccoup}
\end{eqnarray}
where $\hat{\psi}$ denotes the electron field operator, 
and $\langle... \rangle_{n,\lambda}$ indicates 
the vacuum expectation value with respect to the system with the
given density $n_{\sigma}({\bf r})$ and the coupling $\lambda e^2$.
Note that the given density is kept unchanged for any $\lambda$ 
by applying the $\lambda$-dependent external potential.
The EXX energy $E_{\rm x}[n]$ as a functional of the density 
is defined as the $e^2$ term of $E_{\rm xc}[n]$.
As a correlation energy functional $E_{\rm c}[n]$,
we only take the RPA contribution included in the expression of
Eq.(\ref{exccoup}):
\begin{eqnarray}
&&E^{\rm RPA}_{\rm c}[n] =
  \frac{i}{2} \int_0^1 \frac{d \lambda}{\lambda}
   {\rm Tr}[(1- v_{\lambda} D^0 )^{-1} v_{\lambda} D^0
     - v_{\lambda} D^0 ] 
= \frac{-i}{2} {\rm Tr}[\log(1- v D^0 ) + v D^0], \label{excore} \\
&&D^0_\sigma({\bf r}_1,{\bf r}_2,\omega) = 
 \sum^{\rm occ.}_i \sum^{\rm unocc.}_j 
 \psi^{i*}_\sigma({\bf r}_1) \psi^i_\sigma({\bf r}_2)    
 \psi^{j*}_\sigma({\bf r}_2) \psi^j_\sigma({\bf r}_1)  
\Biggl\{ \frac{1}{\omega -\epsilon_\sigma^j +\epsilon_\sigma^i+i \delta}
         -\frac{1}{\omega +\epsilon_\sigma^j -\epsilon_\sigma^i-i \delta}
 \Biggr\},\label{defd0}
\end{eqnarray}
where the trace is taken for index 
$1 \equiv {\bf r} t \sigma$ 
(in Eq.(\ref{excore}), we suppress a factor $1/\int_{-\infty}^\infty dt$ 
for simplicity).
$v_{\lambda}$ is defined as
$v_{\lambda}(1,2)=\lambda e^2 \delta(t_1 - t_2) 
\delta_{\sigma_1 \sigma_2}/|{\bf r}_1-{\bf r}_2| $, 
and $v$ denotes $v_{\lambda=1}$. 
Formally Eq.(\ref{excore}) is the same as
that for the RPA correlation energy of the homogeneous electron gas 
\cite{fetterwalecka} except that $E^{\rm RPA}$ should be treated as 
a functional of density $n_{\sigma}({\bf r})$
through the Lindhard dielectric function $D^0$, 
which is spin-diagonal and constructed from
the eigenfunctions $\psi^i_\sigma({\bf r})$ satisfying  
\begin{eqnarray}
 &&[\frac{-\nabla^2}{2m} + V^{\rm eff}_\sigma({\bf r})-\epsilon^i_\sigma] 
\psi^i_\sigma({\bf r}) = 0, \\
 &&n_{\sigma}({\bf r}) = \sum^{\bf occ.}_i |\psi^i_\sigma({\bf r})|^2.
\end{eqnarray}
Under the assumption of one-to-one correspondence between
$n_{\sigma}({\bf r})$ and $V^{\rm eff}_\sigma({\bf r})$ 
(we define $V^{\rm eff}$ so that the chemical potential
is included in it),
we can treat $\psi^i_\sigma({\bf r})$, and hence $E^{\rm RPA}$,
as a functional of $n_{\sigma}({\bf r})$.
The approximation of Eq.(\ref{excore}) was suggested 
by Gross, Dobson and Petersilka in Ref.\cite{gross96a} (we omit the term 
$f_{\rm xc}$ corresponding to the vertex correction). 

The derivative $\delta E_{\rm c}$ with respect to $\delta D^0$
can be written as
\begin{eqnarray}
 &&\delta E_{\rm c} = \frac{i}{2} {\rm Tr}[W_{\rm p} \delta D^0], 
 \label{deltaec} \\
 &&W_{\rm p} = v_{\rm sc} - v, 
\end{eqnarray}
where $v_{\rm sc} \equiv (1 - v D^0)^{-1} v$ 
denotes the dynamical screened Coulomb interaction in RPA.
We evaluate $\delta E_{\rm c}$ in a static approximation, i.e.,
we replace $W_{\rm p}({\bf r}_1, {\bf r}_2, t_1 - t_2)$ 
with $ W_{\rm p}^{\omega=0}({\bf r}_1, {\bf r}_2)
\times \delta(t_1 -t_2)$, 
where we define $W_{\rm p}^{\omega=0}({\bf r}_1, {\bf r}_2) \equiv
\int_{-\infty}^{\infty} dt W_{\rm p}({\bf r}_1, {\bf r}_2, t)$.
This approximation is justifiable if 
the relaxation time of the dynamical screening $v_{\rm sc}$, 
typically the plasma oscillation time-scale,
is sufficiently shorter than that of the density fluctuation $i D^0$.
We know $i D^0_\sigma({\bf r}_1, {\bf r}_2, t_1=t_2)=
n_\sigma({\bf r}_1)\delta({\bf r}_1-{\bf r}_2)
-[n_\sigma({\bf r}_1, {\bf r}_2)]^2$. 
Here we use the non-local density 
$n_\sigma({\bf r}_1, {\bf r}_2) \equiv \sum^{\rm occ.}_i 
\psi^{i*}_\sigma({\bf r}_1) \psi^i_\sigma({\bf r}_2)$.
Then we obtain
$\delta E_{\rm c} = \delta E_{\rm c1} + \delta E_{\rm c2}$,
where we define
\begin{eqnarray}
&&\delta E_{\rm c1} \equiv 
\frac{-1}{2} \sum_{\sigma} \int d{\bf r}_1 d{\bf r}_2
   W_{\rm p}^{\omega=0}({\bf r}_1, {\bf r}_2) 
  \delta ([n_\sigma({\bf r}_1, {\bf r}_2)]^2),    \label{deltaec1} \\
&&\delta E_{\rm c2} \equiv 
\frac{1}{2} \sum_\sigma \int d{\bf r} W_{\rm p}^{\omega=0}({\bf r}, {\bf r}) 
  \delta n_\sigma({\bf r}).                     \label{deltaec2}
\end{eqnarray}
$\delta E_{\rm c1}$ and $\delta E_{\rm c2}$ correspond to
the correlated part of the screened exchange
and the Coulomb-hole terms, respectively
(see p.40 in Ref.\cite{hedin69}).
As for $\delta E_{\rm c1}$, we can calculate its functional
derivative $\delta E_{\rm c1}/\delta V^{\rm eff}_\sigma({\bf r})$
from $W_{\rm p}^{\omega=0}({\bf r}_1, {\bf r}_2)$,
$n_\sigma({\bf r}_1, {\bf r}_2)$ and
$\delta n_\sigma({\bf r}_1, {\bf r}_2)/
\delta V^{\rm eff}_\sigma({\bf r})$
through Eq.(\ref{deltaec1}). 
Then we obtain 
$\delta E_{\rm c1}/\delta n_{\sigma}({\bf r})$ by 
the same inversion method as was used to obtain 
$\delta E_{\rm x}/\delta n_{\sigma}({\bf r})$ \cite{kotani94} 
(see Eq.(\ref{inv})).
If we evaluate $\delta E_{\rm c}$ from Eq.(\ref{deltaec}) without
the static approximation, we obtain the result essentially
equivalent to Eq.(27) in Ref.\cite{godby88}, 
which was used by Godby, Shl\"uter and Sham
to discuss the eigenvalues of the density functional theory beyond LDA.

In order to calculate all the related quantities,
we exploit the atomic sphere approximation (ASA).
Any points in the space are denoted by $({\bf r},R)$,
where $R$ is the index for atomic sphere (AS)
and ${\bf r}=(r ,\theta, \phi)$
is a vector denoting the position in each AS.
We consider $E_{\rm xc}[n^{\rm s}]$ 
as a functional of $n^{\rm s}_{\sigma}(r,R)$, 
where $n^{\rm s}_{\sigma}(r,R)$  denotes the spherically-averaged density.
$W_{\rm p}^{\omega=0}$ are calculated 
by the product-basis method \cite{aryasetiawan94} as
\begin{eqnarray}
&& W_{\rm p}^{\omega=0}({\bf r}_1,R_1, {\bf r}_2,R_2)
= \sum_{i,j} W_{\rm p}^{\omega=0}(i,R_1,j,R_2)
    \tilde{B}_{i}({\bf r}_1) \tilde{B}_{j}({\bf r}_2),
\end{eqnarray}
where $\tilde{B}_{i}({\bf r})$'s form the product basis 
(see Ref.\cite{aryasetiawan94} for notation). 
In the method of the LMTO-ASA \cite{lmtoref}, the non-local density 
$n_\sigma({\bf r}_1,R_1, {\bf r}_2,R_2)$ is given as a functional
of $V^{\rm eff}_\sigma(r,R)$ through the potential parameters
and radial basis functions. 
Therefore $\delta E_{\rm c1}/\delta V^{\rm eff}_{\sigma}(r,R)$ 
can be evaluated 
by use of the derivative chain rule 
in the same manner used for $E_{\rm x}$ \cite{kotani94}.
Now, we can calculate the contribution to the correlation potential 
$V^{\rm c1}_\sigma(r,R) \equiv 
{\delta E_{\rm c1}[n^{\rm s}]}/{\delta n^{\rm s}_{\sigma}(r,R)}$
by the inversion equation,
\begin{equation}
  \frac{\delta E_{\rm c1}}{\delta V^{\rm eff}_\sigma(r,R)}
  = \sum_{R'}\int_0^{\bar{R'}} dr'
    \frac{\delta n^{\rm s}_{\sigma}(r',R')}{\delta V^{\rm eff}_\sigma(r,R)} \times
    \frac{\delta E_{\rm c1}[n^{\rm s}]}{\delta n^{\rm s}_{\sigma}(r',R')}.
\label{inv}
\end{equation}
Eq.(\ref{inv}) is essentially the same
as the one for $E_{\rm x}$ in Ref. \cite{kotani94}.
Adding spin-independent $V^{\rm c2}(r,R) \equiv$
(the spherically-averaged $W_{\rm p}^{\omega=0}({\bf r},R, {\bf r},R)$), 
we obtain the RPA correlation potential $V^{\rm c}$. 
Since we treat only metals, we can determine constant parts of $V^{\rm x}$ 
and $V^{\rm c1}$ uniquely
(note that $V^{\rm eff}$ includes the chemical potential).
We have developed a code to perform the non-relativistic 
self-consistent calculation with $V^{\rm c}$ 
together with $V^{\rm x}$.
This is done by combining two codes, one calculating $W_{\rm p}$ 
and the other LMTO-ASA EXX code; the former
is a part of the $GW$ program \cite{aryasetiawan92}\cite{aryasetiawan95} 
provided by Aryasetiawan, 
and the latter is developed starting from LMTO-4 \cite{schilf92}.

We show results for Cu(fcc), Ni(fcc), Fe(bcc), and Co(fcc) 
with the lattice constants determined by LDA , 
6.76, 6.55, 5.27,and 6.54 a.u.\cite{mjw}, respectively. 
For the evaluation of $\delta E_{\rm x}$ and $\delta E_{\rm c}$,
we take all the pairs $(R_1, R_2)$ 
within the second nearest neighbors
and treat $4s$, $4p$, and $3d$ as valence orbitals. 
For the calculation of $W_{\rm p}^{\omega=0}$, 
we use 35-$k$ points in the irreducible Brillouin zone,
with the Lorenzian broadening $\delta=0.02$ Ry. 
in Eq.(\ref{defd0}). 
We used 90 optimum product basis $B_i$'s, where
all the core eigenfunctions are used to generate them.

In Fig.\ref{vxcfig}, we show the self-consistent 
$V^{\rm xc}= V^{\rm x}+ V^{\rm c1} + V^{\rm c2}$ for Cu, Ni and Fe.
Constant parts of $V^{\rm x}$ and $V^{\rm c1}$
are properly fixed for presentation.
We find that the large dips of $V^{\rm x}$ and 
the large differences between up- and down-spin components of $V^{\rm x}$
are strongly reduced by adding $V^{\rm c}$. 
As a result the final $V^{\rm xc}$ obtained by EXX+RPA 
gets much closer to those 
by LDA \cite{barth72} than it was for $V^{\rm x}$ obtained by EXX. 
$V^{\rm c2}$, which corresponds to the screening length \cite{vc2note},
has considerable structures but its contribution to $V^{\rm c}$
is largely cancelled by $V^{\rm c1}$ in the vicinity of the core region.
This can be explained by the fact that 
$\delta n({\rm r}_1,{\rm r}_2)$ is sufficiently short-ranged
for given $\delta n({\rm r})$ near the core region, 
which allows us to evaluate $\delta E_{\rm c}$ approximately
by use of 
$W_{\rm p}^{\omega=0}({\rm r}_1,{\rm r}_1)$
instead of $W_{\rm p}^{\omega=0}({\rm r}_1,{\rm r}_2)$
in Eqs.(\ref{deltaec1}-\ref{deltaec2}):
By this replacement, 
we have $\delta E_{\rm c1}+\delta E_{\rm c2}=0$ because of
$\int d{\bf r}_1 D^0_\sigma({\bf r}_1, {\bf r}_2, t_1=t_2)=0$.
On the other hand, we see a rather large cancellation between
$V^{\rm x}$ and $V^{\rm c1}$ out of the core region ($r \gtrsim$ 1.0 a.u.).
This is because the main contribution to 
$\delta E_{\rm c1}$ in Eq.(\ref{deltaec1}) 
comes from the integral out of the range of
$v_{sc}^{\omega=0}$, where we expect the behavior
of $W_{\rm p}^{\omega=0}$ is $W_{\rm p}^{\omega=0}({\rm r}_1,{\rm r}_2)
\sim -1/|{\rm r}_1-{\rm r}_2|$.

The eigenvalue dispersions are shown in Fig.\ref{bandfig}
together with those calculated by usual LDA, and EXX. 
Results by EXX+RPA are very close to those obtained by LDA, 
and are much different from the results by EXX.
The magnetic moments shown in Table \ref{magnet} 
are also in good agreement with the LDA's and experiments.
To obtain numerically accurate values of the magnetic moments,
it is necessary to extrapolate the series of 
fully converged (about the number of $k$ points)
results for different $\delta$ towards $\delta=0$. 
This, however, seems too time-consuming in the present stage 
of our unoptimized computer code, and we only show the results 
for $\delta=0.04$ Ry. with 35-$k$ points 
in Table \ref{magnet} for reference.

Let us discuss some possible sources of the
difference between our results and those of the true DF. These are
(1) the product-basis expansion,
(2) RPA, and (3) static approximation for RPA.
As for (1), our test for a model interaction
$v_{\rm sc}=\exp(-\kappa r)/r$ (in this case we can obtain 
accurate results without using the product-basis expansion) indicates that
the potential $V_{\rm c}$ is poorer when getting closer to the AS's center. 
The oscillation of $V_{\rm c}$ near the nucleus seems to be an artifact
due to the method of expansion. 
However, this fortunately little affects on the energy bands 
and the magnetic moments (we will show the details in the subsequent paper).
To take into account the effects beyond RPA, (2), we have to treat $f_{\rm xc}$
in Ref. \cite{gross96a}, which corresponds to the vertex correction. 
Some parts of the contribution due to $f_{\rm xc}$ might be taken into account
by LDA-like approximations; one of the simple ways is to add
the difference between the LDA XC calculated by RPA \cite{barth72} 
and the one calculated by a more accurate scheme \cite{vosko80}. 
We tried the above method but the correction turned out to be rather small;
the magnetization of Fe enhanced by 0.05 $\mu_{\rm B}$ 
(this value is similar to the corresponding LDA case).
However, we are not so confident whether the correction is really 
meaningful or not. 
As for (3), the dynamical effects will be simulated in effect by
making $v_{\rm sc}^{\omega=0}$ closer to $v$ (no relaxation limit). 
It should reduce the magnitude of $V_{\rm c}$. 
Therefore the position of $d$ band relative to $s$ band should be
somehow pushed down for Cu case (See Fig.\ref{vxcfig}). 
Contrary, this estimation concerning (3) is opposite in the case of LDA.
In the case of LDA, we can rather easily show that 
$d$ bands calculated by LDA in the RPA \cite{barth72}
are pushed up from those obtained by its static approximation.
In conclusion, it seems rather difficult to evaluate (3) 
based on the LDA idea; we have not succeeded in giving any 
reasonable evaluations for the magnitude of errors due to (2) and (3).

It sometimes have been claimed that the itinerant-magnetism 
is a sensitive problem where the higher-order electron-correlation  
\cite{kanamori63} beyond RPA screening might be essential 
in determining the magnetic moments or the ground states.
Our results, however, indicate that the essential part 
determining the magnetic moments might be well accounted for
by the correlation within the RPA screening,
even though the higher-order correlations are certainly important
for excitations as was discussed in Ref. \cite{aryasetiawan92}.
Incidentally, as for the the higher-order correlations,
we as well can extend the OEP method
to include other classes of correlation diagrams 
included in Eq.(\ref{exccoup})
and to evaluate their contributions in an ab-initio way.
For example, the exchange diagrams of order $e^4$ 
that is important to cancel the self-interaction of order $e^4$
can be included.
Further, the ladder types of diagrams that include 
the free energy due to the spin wave 
should be taken into account in the case of finite temperatures. 

Our method determines not only $V^{\rm eff}$, but also
$W_{\rm p}^{\omega=0}$ in a self-consistent manner.
$W_{\rm p}^{\omega=0}$ calculated from the LDA
eigenfunctions (and eigenfunctions) seems reasonable for transition metals: 
For example, the magnetic moments 2.07 $\mu_B$ of Fe, 
which is calculated with a fixed $W_{\rm p}^{\omega=0}$ 
calculated from the LDA eigenvalues,
is very close to the self-consistent value 2.04 $\mu_B$ 
in Table \ref{magnet}.
However, such results obtained by the LDA eigenvalues
are not expected to be sufficient for solids like transition-metal oxides.
In Ref.\cite{aryasetiawan95} for NiO, it is shown that 
the $GW$ calculation with $W_{\rm p}$ constructed from
the LDA eigenvalues gives poor results
because of the too small a LDA band gap, and that
the self-consistency of $GW$ is essentially needed.
The fully self-consist calculation of $GW$ seems
very hard mainly due to the programing difficulties
and the lack in computational power. 
From such a point of views, our self-consistent calculation 
might be a good substitution of the self-consistent $GW$.

We have not developed the code to calculate $E_{\rm c}^{\rm RPA}$ 
itself yet. We only treated its functional derivative
$\delta E_{\rm c}^{\rm RPA}/\delta n^{\rm s}$, 
and the static approximation is not applicable 
for the total energy calculation.
To our knowledge, none have done the ab-initio
calculation of $E_{\rm c}^{\rm RPA}$. 
It requires to treat unoccupied states precisely.
It might be a little difficult in LMTO-ASA scheme.
However, we have to try such calculations for
determining lattice constants and other cohesive properties.

I thank F. Aryasetiawan for his $GW$ program and for
stimulating discussions, and M. van Schilfgaarde, 
T. A. Paxton, O. Jepsen, and O. K. Andersen
for their TB-LMTO program.
I also thank H. Akai for discussions and 
for critical reading of the manuscript.

\newpage
\narrowtext

\newpage
\begin{figure}
\caption[]
 {Exact exchange potential $V^{\rm x}$ and RPA correlation potential 
  $V^{\rm c}=V^{\rm c1}+V^{\rm c2}$ for Cu, Ni and Fe. 
  The LDA XC potential \cite{barth72} 
  as references are calculated for the density 
  determined by the self-consistent calculation of EXX+RPA.} 
\label{vxcfig}
\end{figure}

\begin{figure} 
 \caption[]{ Energy dispersion curves calculated 
 by three different types of self-consistent calculations, 
 LDA, EXX, and EXX+RPA. EXX denotes the calculation
 which is performed with LDA correlation\cite{barth72} potential
 in addition to the EXX potential.}
\label{bandfig}
\end{figure}

\newpage
\begin{table}
\caption[]
  {Calculated spin magnetic moments ($\mu_{\rm B}$). 
   For EXX+RPA, we show the values obtained 
   with $\delta=0.02$ Ry, and $\delta=0.04$ Ry (parenthesis), 
   see Eq.(\ref{defd0}). 
   The lattice constants used in the calculations are somehow smaller 
   than the experimental ones (see text). 
   Experimental spin magnetic moments are deduced from the experimental
   values of the saturated magnetization and g-values \cite{reck69}.}
\begin{tabular}{ccccc}
    &  LDA  & EXX   & EXX+RPA      & Expt. \\ \tableline
 Fe &  2.13 & 3.27  & 2.04 (2.07)  & 2.12  \\
 Co &  1.54 & 2.29  & 1.51 (1.52)  & 1.59  \\
 Ni &  0.58 & 0.68  & 0.57 (0.58)  & 0.56  \\
\end{tabular}
\label{magnet}
\end{table}


\begin{references}
\bibitem{talman76} J. D. Talman and W. F. Shadwick,
  Phys. Rev. A{\bf 14}, 36 (1976).
\bibitem{kotani94}
  T. Kotani, Phys. Rev. Lett. {\bf 74} 2989 (1995);
  Phys. Rev. B {\bf 50} 14816 (1994), 
  {\it ibid.} {\bf 51} 13903(E) (1995);
  T. Kotani and H. Akai, {\it ibid.} {\bf 52} 17153 (1995);
\bibitem{kotani96fe}
  T. Kotani and H. Akai, Physica B in press.
\bibitem{kotani96kkr}
  T. Kotani and H. Akai, Phys. Rev. B {\bf 54} 16502 (1996).
\bibitem{aashamar79} K. Aashamar, T. M. Luke, and J. D. Talman,
  J. Phys. B{\bf 12}, 3455 (1979).
\bibitem{garbo95}
  T. Garbo, and E. K. U. Gross,
  Chem. Phys. Lett. {\bf 240}, 141 (1995);
  to appear in Int. J. Qua. Chem. (1997).
\bibitem{aryasetiawan94} 
  F. Aryasetiawan and O. Gunnarson,
  Phys. Rev. B {\bf 49}, 16214 (1994).
\bibitem{fetterwalecka} 
 See, e.g, A. L. Fetter and J. D. Walecka, 
 {\it Quantum Theory of Many-Particle Systems},
 (McGraw-Hill, New York, 1971).
\bibitem{gross96a}
 E.K.U.Gross, J.F. Dobson, and M. Petersilka, in
  {\it Density functional theory}, edited by
  R.F. Nalewajski, Topics in Current Chemistry,  
  (Springer-Verlag, Berlin Heidelberg, 1996) Vol. 181, p. 81.
\bibitem{hedin69}
  L. Hedin and S. Lundqvist, in 
  {\it Solid State Physics}, edited by
  H. Ehnrenreich, F. Seitz, and D. Turnbull 
  (Academic, New York, 1969), Vol. 23, p. 1.
\bibitem{godby88}
  R. W. Godby, M. Shl\"uter, and L. J. Sham,
  Phys. Rev. B {\bf 37}, 10159 (1988);
\bibitem{lmtoref}
 O. K. Andersen, O. Jepsen, and D. Gl\"otzel, 
 in {\it Highlights of Condensed-Matter
 Theory}, edited by F.Bassani, F. Fumi, and M. P. Tosi 
 (North-Holland, Amsterdam, 1985), p.59;
\bibitem{aryasetiawan92} 
  F. Aryasetiawan,
  Phys. Rev. B {\bf 46}, 13051 (1992);
\bibitem{aryasetiawan95} 
  F. Aryasetiawan and O. Gunnarson,
  Phys. Rev. Lett {\bf 74}, 3221 (1995),
  F. Aryasetiawan and K. Karlsson,
  Phys. Rev. B {\bf 54}, 5353 (1996)
\bibitem{schilf92}
  The TB-LMTO program ver. 4, 
  by M. van Schilfgaarde, T. A. Paxton, O. Jepsen, 
  and O. K. Andersen, 
  Max-Planck-Institut f\"ur Festk\"orperforschung, 
  Federal Republic of Germany, (1992)
\bibitem{mjw}
  V. L. Moruzzi, J. F. Janak, and A. R. Williams,
 {\it Calculated Electronic Properties of Metals},
 (Pergamon press inc., New York, 1971).
\bibitem{barth72}
 The RPA-level of LDA by 
 U. von Barth and L. Hedin, J. Phys. C{\bf 5}, 1629 (1972).
\bibitem{vc2note}
 If we assume a simple-form for the screening, i.e.,
 $v_{sc}^{\omega=0}({\bf r}_1,{\bf r}_2)=
 \exp(-\kappa |{\bf r}_1-{\bf r}_2|)/|{\bf r}_1-{\bf r}_2|$,
 we can set $V^{\rm c2} = -\kappa$ following its definition.
\bibitem{vosko80}
 S. H. Vosko, L. Wilk, and M Nusair,
 Can. J. Phys. {\bf 58}, 1200, (1980)
\bibitem{kanamori63}
  J. Kanamori, Prog. Theor. Phys. {\bf 30}, 275 (1963)   
\bibitem{reck69}
  R. A. Reck and D. J.Fry, Phys. Rev. {\bf 184}, 492 (1969);
  H. Denan, A. Herr, and A. J. P. Meyer, Jour. App. Phys. {\bf 39}, 69 (1968);
  M. J. Besnus, A. J. P. Meyer, and R. Berninger, Phys. Lett. {\bf 32}A, 
  192 (1970)
\end{references}
\end{document}